\documentclass[a4paper,UKenglish,cleveref, autoref, thm-restate]{lipics-v2021}

\nolinenumbers
\title{Simplified Tight Bounds for Monotone Minimal Perfect Hashing} 

\author{Dmitry Kosolobov}{Institute of Natural Sciences and Mathematics, Ural Federal University, Ekaterinburg, Russia}{dkosolobov@mail.ru}{https://orcid.org/0000-0002-2909-2952}{}

\authorrunning{D. Kosolobov}

\Copyright{Dmitry Kosolobov}



\category{} 

\relatedversion{} 

\funding{\phantom{This work was carried out with financial support from the Ministry of Science and Higher Education of the Russian Federation (project 075-02-2024-1428 for the development of the regional scientific and educational mathematical center ``Ural Mathematical Center'').}}

\hideLIPIcs  

\EventEditors{Shunsuke Inenaga and Simon J. Puglisi}
\EventNoEds{2}
\EventLongTitle{35th Annual Symposium on Combinatorial Pattern Matching (CPM 2024)}
\EventShortTitle{CPM 2024}
\EventAcronym{CPM}
\EventYear{2024}
\EventDate{June 25--27, 2024}
\EventLocation{Fukuoka, Japan}
\EventLogo{}
\SeriesVolume{296}
\ArticleNo{6}

\begin{document}

\maketitle

\begin{abstract}
	Given an increasing sequence of integers $x_1,\ldots,x_n$ from a universe $\{0,\ldots,u-1\}$, the monotone minimal perfect hash function (MMPHF) for this sequence is a data structure that answers the following rank queries: $rank(x) = i$ if $x = x_i$, for $i\in \{1,\ldots,n\}$, and $rank(x)$ is arbitrary otherwise. Assadi, Farach-Colton, and Kuszmaul recently presented at SODA'23 a proof of the lower bound $\Omega(n \min\{\log\log\log u, \log n\})$ for the bits of space required by MMPHF, provided $u \ge n 2^{2^{\sqrt{\log\log n}}}$, which is tight since there is a data structure for MMPHF that attains this space bound (and answers the queries in $O(\log u)$ time). In this paper, we close the remaining gap by proving that, for $u \ge (1+\epsilon)n$, where $\epsilon > 0$ is any constant, the tight lower bound is $\Omega(n \min\{\log\log\log \frac{u}{n}, \log n\})$, which is also attainable; we observe that, for all reasonable cases when $n < u < (1+\epsilon)n$, known facts imply tight bounds, which virtually settles the problem. Along the way we substantially simplify the proof of Assadi et al.~replacing a part of their heavy combinatorial machinery by trivial observations. However, an important part of the proof still remains complicated. This part of our paper repeats arguments of Assadi et al.~and is not novel. Nevertheless, we include it, for completeness, offering a somewhat different perspective on these arguments.
\end{abstract}

\section{Introduction}
\label{sec:intro}

The \emph{monotone minimal perfect hash function (MMPHF)} is a data structure built on an increasing sequence $x_1 < \cdots < x_n$ of integers from a universe $\{0,\ldots,u-1\}$ that answers the following \emph{rank} queries: $rank(x) = i$ if $x = x_i$ for some $i$, and $rank(x)$ is arbitrary otherwise. 

The MMPHF is an important basic building block for succinct data structures (e.g., see~\cite{Belazzougui2,BelazzouguiNavarro,BelazzouguiCunialKarkkainenMakinen,GrossiOrlandiRaman,GagieNavarroPrezza2,CliffordEtAl}). It turns out that the relaxation that permits to return arbitrary answers when $x$ does not belong to the stored sequence leads to substantial memory savings: as was shown by Belazzougui, Boldi, Pagh, and Vigna~\cite{BelazzouguiBoldiPaghVigna}, it is possible to construct an MMPHF that occupies $O(n \min\{\log\log\log u, \log n\})$ bits of space with $O(\log u)$-time queries, which is a remarkable improvement over the $\Omega(n\log\frac{u}{n})$ bits required to store the sequence $x_1,\ldots,x_n$ itself.

Until very recently, the best known lower bound for the space of the MMPHF was $\Omega(n)$ bits, which followed from the same bound for the minimal perfect hashing (see~\cite{FredmanKomlos,Mehlhorn,Radhakrishnan}). In 2023, this bound was improved by Assadi, Farach-Colton, and Kuszmaul \cite{AssadiEtAl}: they proved that, surprisingly, the strange space upper bound $O(n \min\{\log\log\log u, \log n\})$ is actually tight, provided $u \ge n 2^{2^{\sqrt{\log\log n}}}$. Thus, the problem was fully settle for almost all possible $u$. Their proof utilized a whole spectre of sophisticated combinatorial techniques: a ``conflict graph'' of possible data structures, the fractional chromatic number for this graph, the duality of linear programming, non-standard graph products, and intricate probabilistic arguments. 

In this paper we simplify their proof, removing all mentioned concepts except the intricate probabilistic arguments, and we slightly extend the result: our lower bound is $\Omega(n \min\{\log\log\log\frac{u}{n}, \log n\})$, where $u \ge (1 + \epsilon)n$ for an arbitrary constant $\epsilon > 0$ (the $\Omega$ hides an $\epsilon\log\frac{1}{\epsilon}$ factor). Further, we show that this bound is tight by devising a simple extension of the MMPHF data structure of Belazzougui et al.~\cite{BelazzouguiBoldiPaghVigna}. We also observe that, for all reasonable cases $n < u < (1 + \epsilon)n$, tight bounds can be obtained using known facts.

All ingredients required for our simplification were actually already present in the paper by Assadi et al. For instance, we eventually resort to essentially the same problem of coloring random sequences (see below) and the same probabilistic reasoning. Our extension of the MMPHF by Belazzougui et al.~\cite{BelazzouguiBoldiPaghVigna} is also not difficult. It seems that the simpler proof and the extension were overlooked.

The paper is organized as follows. In Section~\ref{sec:prelim}, we define our notation and discuss weaker lower bounds and tight upper bounds, including our extension of the MMPHF from~\cite{BelazzouguiBoldiPaghVigna}. Section~\ref{sec:colorings} provides a short way to connect the lower bound to certain colorings of the universe. Section~\ref{sec:to-large-universe} shows how the problem can be further reduced to the coloring of certain random sequences on a very large universe $u = 2^{2^{n^3}}$; Assadi et al.~make a similar reduction but their explanation includes unnecessary non-standard graph products and does not cover the extended range $(1+\epsilon)n \le u$. The material in all these first sections is quite easy. Finally, Section~\ref{sec:large-universes} is a core of the proof, which, unfortunately, still involves complicated arguments. This part is exactly equivalent to Lemma 4.2 in~\cite{AssadiEtAl}, which was also the most challenging part in~\cite{AssadiEtAl}. For the self-containment of the paper, instead of directly citing Lemma 4.2 from~\cite{AssadiEtAl}, we decided to offer a somewhat different view on the same arguments. Depending on their disposition, the reader might find our exposition more preferrable or vice versa; it shares the central idea with~\cite{AssadiEtAl} but reaches the goal through a slightly different path.

\section{Tight Upper Bounds}
\label{sec:prelim}

Denote $[p..q] = \{k \in \mathbb{Z} \colon p\le k\le q\}$, $[p..q) = [p..q{-}1]$, $(p..q] = [p{+}1..q]$, $(p..q) = [p{+}1..q{-}1]$. Throughout the text, $u$ denotes the size of the universe $[0..u)$, from which the hashed sequence $x_1,\ldots,x_n$ is sampled, and $n$ denotes the size of this sequence. All logarithms have base $2$. To simplify the notation, we assume that $\log\log\log\frac{u}{n} = \Omega(1)$, for any $\frac{u}{n} > 0$.

Let us overview known upper bounds for the space required by the MMPHF. The MMPHF of Belazzougui et al.~\cite{BelazzouguiBoldiPaghVigna} offers $O(n\log\log\log u)$ bits of space. When $n 2^{2^{\sqrt{\log\log n}}} \le u \le 2^{2^{\mathrm{poly}(n)}}$, it is tight due to the lower bound of~\cite{AssadiEtAl}. For larger $u$, we can construct a perfect hash $h:[0..u) \to [1..n]$ that occupies $O(n\log n)$ bits and bijectively maps the hashed sequence $x_1,\ldots,x_n$ onto $[1..n]$ (e.g., it might be the classical two-level scheme~\cite{FredmanKomlosSzemeredi} or a more advanced hash~\cite{BelazzouguiBotelhoDietzfelbinger}) and we store an array $A[1..n]$ such that, for $i \in [1..n]$, $A[h(x_i)]$ is the rank of $x_i$. This scheme takes $O(n\log n)$ bits, which is tight when $u \ge 2^{2^{\mathrm{poly}(n)}}$, again due to the lower bound of~\cite{AssadiEtAl} since $n\log\log\log 2^{2^{\mathrm{poly}(n)}} = \Theta(n\log n)$. For small $u$ (like $u = \Theta(n)$), we can simply store a bit array $B[0..u{-}1]$ such that $B[x] = 1$ iff $x = x_i$. Such ``data structure'' takes $u$ bits and can answer the rank queries for the MMPHF (very slowly). With additional $o(n)$ bits \cite{Clark,Jacobson}, one can answer in $O(1)$ time the rank queries on this array and, also, the following \emph{select queries}: given $i \in [1..n]$, return the position of the $i$th $1$ in the array.

The described MMPHFs give the tight space upper bound $O(n \min\{\log\log\log u,\log n\})$, for $u \ge n 2^{2^{\sqrt{\log\log n}}}$, and an upper bound $O(n)$, for $u = O(n)$. Let us construct an MMPHF that occupies $O(n \log\log\log\frac{u}{n})$ bits, for $2n \le u < n 2^{2^{\sqrt{\log\log n}}}$, which, as we show below, is tight. (We note that a construction similar to ours was alluded in~\cite{BelazzouguiNavarro,BelazzouguiNavarro2}.) We split the range $[0..u)$ into $n$ buckets of length $b = \frac{u}{n}$. For $i \in [1..n]$, denote by $n_i$ the number of elements of the sequence $x_1,\ldots,x_n$ that are contained in the $i$th bucket. We construct the MMPHF of Belazzougui et al.~for each bucket, thus consuming $O(\sum_{i=1}^n n_i \log\log\log\frac{u}{n}) = O(n\log\log\log\frac{u}{n})$ bits of space. Then, we build a bit array $B[1..2n]$ such that $B[\sum_{j=1}^{i-1} n_j + i] = 1$, for each $i\in [1..n]$, and all other bits are zeros; it is convenient to view $B$ as the concatenation of bit strings $10^{n_i}$, for $i \in [1..n]$, where $0^{n_i}$ denotes a bit string with $n_i$ zeros. The bit array $B$ supports select queries and takes $O(n)$ bits. To answer the rank query for $x \in [0..u)$, our MMPHF first calculates $i = \lfloor x / b\rfloor+1$ (the index of the bucket containing $x$), then it computes the position $k$ of the $i$th $1$ in the bit array $B$ using the select query, and the answer is equal to $k - i$ plus the answer of the query $rank(x - (i-1) b)$ in the MMPHF associated with the $i$th bucket. We, however, cannot afford to store $n$ pointers to the MMPHFs associated with the buckets. Instead, we concatenate the bit representations of these MMPHFs and construct another bit array $N$ of length $O(n\log\log\log\frac{u}{n})$ where the beginning of each MMPHF in the concatenation is marked by $1$ (and all other bits are zeros). The array $N$ also supports the select queries and the navigation to the MMPHF associated with the $i$th bucket is performed by finding the $i$th $1$ in the array $N$ using the select query.

With the described data structures, we obtain tight upper bounds for all $u \ge (1 + \epsilon)n$, where $\epsilon > 0$ is an arbitrary constant. In particular, when $(1+\epsilon)n \le u < 2n$, the upper bound $O(u)$ is equal to $O(n)$ and it is known to be tight: an analysis of the minimal perfect hashing \cite{Mehlhorn,Radhakrishnan} implies the lower bound $\Omega(n)$ for the MMPHF, provided $u \ge (1 + \epsilon)n$. More precisely, the bound is $(u-n)\log\frac{u}{u-n} - O(\log n)$, which holds for arbitrary $u > n$ (see~\cite{BelazzouguiBotelhoDietzfelbinger,Mehlhorn,Radhakrishnan}). For illustrative purposes, we rederive this bound in a proof sketch provided in Section~\ref{sec:to-large-universe}. 

It remains to analyse the range $n < u < (1 + \epsilon)n$. To this end, we restrict our attention only to matching upper and lower bounds that are greater than $\Omega(\log n)$. This is a reasonable restriction because all these data structures are usually implemented on the word RAM model, where it is always assumed that $\Theta(\log n)$ bits are available for usage. Thus, we analyse only the first term of the difference $(u-n)\log\frac{u}{u-n} - O(\log n)$ assuming that it is at least twice greater than the $O(\log n)$ term. Suppose that $u = (1 + \alpha)n$, where $0 < \alpha < \frac{1}4$ and $\alpha$ is not necessarily constant. Then, we obtain $(u-n)\log\frac{u}{u-n} = n \alpha \log\frac{1+\alpha}{\alpha} = \Theta(n\alpha \log\frac{1}{\alpha})$. It is known that the bit array $B[0..u{-}1]$ such that $B[x] = 1$, for $x=x_i$, can be encoded into $\log\binom{u}{n}$ bits, which can be treated as an MMPHF for the sequence $x_1,\ldots,x_n$. By the well-known entropy inequality~\cite{CoverThomas}, we obtain $\log\binom{u}{n} \le n\log\frac{u}{n} + (u - n)\log\frac{u}{u-n} = n\log(1 + \alpha) + n\alpha\log\frac{1+\alpha}{\alpha} \le O(n\alpha + n\alpha\log\frac{1}{\alpha}) = O(n\alpha\log\frac{1}{\alpha})$, which coincides with the lower bound $(u-n)\log\frac{u}{u-n} - O(\log n) = \Omega(n\alpha\log\frac{1}{\alpha})$ (rederived in Section~\ref{sec:to-large-universe}) and, thus, is tight. In particular, for constant $\alpha = \epsilon$, we obtain the bound $\Omega(n)$ that hides $\epsilon\log\frac{1}{\epsilon}$ under the $\Omega$.

Hereafter, we assume that all presented results hold for sufficiently large $u$. We will mostly consider the case $u \le 2^{2^{\mathrm{poly}(n)}}$, which also implies sufficiently large $n$.

\section{From Data Structures to Colorings}
\label{sec:colorings}

Let us first consider deterministic MMPHFs. Randomized MMPHFs are briefly discussed in Remark~\ref{rem:random} in the end of Section~\ref{sec:to-large-universe}.

Given positive integers $u$ and $n < u$, the MMPHF that uses $S$ bits of space is a data structure that can encode any increasing sequence $x_1 < \cdots < x_n$ from $[0..u)$ into $S$ bits to support the rank queries: for $x \in [0..u)$, $rank(x) = i$ if $x = x_i$ for some $i \in [1..n]$, and $rank(x)$ is arbitrary otherwise. We assume a very powerful model of computation: the query algorithm has unbounded computational capabilities and has unrestricted access to its $S$ bits of memory. Formally, it can be modelled as a function $rank\colon [0..u) \times \{0,1\}^S \to [1..n]$ that takes as its arguments an integer $x\in [0..u)$ and the content of the $S$-bit memory and outputs the rank of $x$ in a sequence encoded in these $S$ bits (note that the same bits might correctly encode many different sequences); it is guaranteed that any increasing sequence $x_1, \ldots, x_n$ has at least one encoding $c \in \{0,1\}^S$ that provides correct queries for it, i.e., $rank(x_i, c) = i$, for $i \in [1..n]$. Our goal is to prove that such function can exist only if $S \ge \Omega(n\log\log\log\frac{u}{n})$, provided $(1+\epsilon)n \le u \le 2^{2^{\mathrm{poly}(n)}}$, for constant $\epsilon > 0$.


The function $rank\colon [0..u) \times \{0,1\}^S \to [1..n]$ can be viewed as a family of $2^S$ colorings of the range $[0..u)$: each ``memory content'' $c \in \{0,1\}^S$ colors any $x \in [0..u)$ into the color $rank(x,c)$, one of $n$ colors $[1..n]$. We say that such a coloring \emph{encodes} a sequence $x_1 < \cdots < x_n$ if the color of $x_i$ is $i$, for $i \in [1..n]$. Note that one coloring may encode many distinct sequences and one sequence may be encoded by different colorings of $[0..u)$.

\begin{example}
	For $u = 17$, the following coloring of $[0..u)$ (the colors are both highlighted and denoted by indices $1$--$5$ below) encodes the sequences $3, 6, 7, 10, 14$, and $1,2,4,9,12$, and $1,6,11,15,16$, to name a few:

\begin{tabular}{ccccccccccccccccc}
	{\color{red}0} & {\color{red}1} & 2 & {\color{red}3} & {\color{blue}4} & {\color{blue}5} & 6 & {\color{blue}7} & {\color{red}8} & {\color{green}9} & {\color{green}10} & {\color{blue}11} & {\color{brown}12} & 13 & {\color{brown}14} & {\color{green}15} & {\color{brown}16} \\
	{\tiny 1} &{\tiny 1} & {\tiny 2} & {\tiny 1} & {\tiny 3} & {\tiny 3} & {\tiny 2} & {\tiny 3} & {\tiny 1} & {\tiny 4} & {\tiny 4} & {\tiny 3} & {\tiny 5} & {\tiny 2} & {\tiny 5} & {\tiny 4} & {\tiny 5} 
\end{tabular}
\end{example}

We thus have deduced that the MMPHF provides a family of $2^S$ colorings that encode all possible sequences of size $n$ from $[0..u)$. Now, if we prove that any such all-encoding family must have at least $C$ colorings, then we will have $2^S \ge C$, which implies the space lower bound $S \ge \log C$. In what follows, we show that $C \ge (\log\log \frac{u}{n})^{\Omega(n)}$ when $(1+\epsilon)n \le u \le 2^{2^{\mathrm{poly}(n)}}$, hence proving the lower bound $S \ge \Omega(n\log\log\log\frac{u}{n})$.

\section{Coloring of Random Sequences}
\label{sec:to-large-universe}

As in~\cite{AssadiEtAl}, we utilize the following probabilistic argument. Consider a random process that generates size-$n$ sequences from $[0..u)$ in such a way that any fixed coloring of $[0..u)$ encodes the generated sequence with probability at most $1 / C$. Now if a family of colorings of $[0..u)$ is such that any size-$n$ sequence from $[0..u)$ can be encoded by some of its colorings (i.e., it is an ``all-encoding'' family as the one provided by the MMPHF), then it necessarily contains at least $C$ colorings since any sequence generated by our random process is encoded with probability~$1$ by one of the colorings. Let us illustrate this reasoning by sketching a proof of a weaker lower bound for our problem (which can also serve as a proof of the space lower bound for the minimal perfect hash function on size-$n$ sequences from $[0..u)$).

Consider a process that generates all size-$n$ sequences from $[0..u)$ uniformly at random. Fix an arbitrary coloring of $[0..u)$ with colors $[1..n]$. Denote by $c_i$ the number of elements $x\in [0..u)$ with color $i$. Since the coloring might encode at most $c_1  \cdots c_n$ distinct size-$n$ sequences from $[0..u)$, the probability that a random sequence is encoded by it is at most $c_1\cdots c_n / \binom{u}{n}$. Since $\sum_{i=1}^n c_i = u$, the maximum of $c_1\cdots c_n$ is attained when all $c_i$ are equal, so $c_1\cdots c_n \le (\frac{u}{n})^n$.  
Thus, we obtain $c_1\cdots c_n / \binom{u}{n} \le (\frac{u}{n})^n / \binom{u}{n}$ and, hence, any ``all-encoding'' family must contain at least $\binom{u}{n} / (\frac{u}{n})^n$ colorings, which, after applying the logarithm, implies the space lower bound $\log (\binom{u}{n} / (\frac{u}{n})^n)$ for the MMPHF. Finally, the entropy inequality~\cite{CoverThomas} $\log\binom{u}{n} \ge n\log\frac{u}{n} + (u-n)\log\frac{u}{u-n} - O(\log n)$ gives the lower bound $(u-n)\log\frac{u}{u-n} - O(\log n)$, which is bounded by $\Omega(n\epsilon\log\frac{1}{\epsilon}) = \Omega(n)$ when $(1+\epsilon)n \le u$ for constant $\epsilon > 0$.

It is evident from this sketch that our random process must be more elaborate than a simple random pick. 

In what follows we essentially repeat the scheme from~\cite{AssadiEtAl}. Namely, for the special case $u = 2^{2^{n^3}}$, we devise a random process generating size-$n$ sequences from $[0..u)$ such that any fixed coloring encodes its generated sequence with probability at most $1 / n^{\Omega(n)}$. Hence, the number of colorings in the family provided by the MMPHF is at least $n^{\Omega(n)}$, which, after applying the logarithm, implies the space lower bound $\Omega(n\log n) = \Omega(n\log\log\log\frac{u}{n})$. All other possible $u$ are reduced to this special case. Let us start with this reduction.

\subparagraph{\boldmath Reduction of arbitrary $u$ to very large $u$.} Suppose that, for any $n$ and $u = 2^{2^{n^3}}$, we are able to devise a random process that generates size-$n$ sequences from $[0..u)$ in such a way that any coloring of $[0..u)$ encodes the generated sequence with probability at most $1 / n^{\Omega(n)}$. Now let us fix arbitrary $u$ and $n$ such that $(1 + \epsilon)n \le u \le 2^{2^{\mathrm{poly}(n)}}$, for constant $\epsilon > 0$. 

Case (i) $u \ge 2^{2^{n^3}}$. For this case, the same random process that generates size-$n$ sequences from $[0..2^{2^{n^3}}) \subseteq [0..u)$ gives the probability at most $1 / n^{\Omega(n)}$, again implying the space lower bound $\Omega(n\log n)$ as above, which is equal to $\Omega(n\log\log\log\frac{u}{n})$ when $2^{2^{n^3}} \le u \le 2^{2^{\mathrm{poly}(n)}}$.

Case (ii) $(1 + \epsilon)n \le u < 2^{2^8}n$. Since $n\log\log\log\frac{u}{n} = \Theta(n)$, the lower bound $\Omega(n)$ for this case was obtained above.

Case (iii) $2^{2^8}n \le u < 2^{2^{n^3}}$. The key observation is that while our hypothesised random process cannot be applied to generate sequences of size $n$ (since $u$ is too small), it can generate smaller sequences, for instance, of size $\bar{n} \le (\log\log u)^{1/3}$ (since $2^{2^{\bar{n}^3}} \le u$). Accordingly, we compose a random process that generates a size-$n$ sequence as follows: it splits the range $[0..u)$ into $n/\bar{n}$ equal blocks of length $\bar{u} = u / (n / \bar{n})$ and independently generates a size-$\bar{n}$ sequence inside the first block, a size-$\bar{n}$ sequence inside the second block, etc. The generation inside each block is performed using our hypothesised random process, which is possible provided $\bar{u} \ge 2^{2^{\bar{n}^3}}$. This inquality is satisfied by putting $\bar{n} = \lfloor(\log\log \frac{u}{n})^{1/3}\rfloor$ (note that $\bar{n} \ge 2$ since $\frac{u}{n} \ge 2^{2^8}$): $\bar{u} \ge u/n = 2^{2^{\log\log\frac{u}{n}}} \ge 2^{2^{\bar{n}^3}}$.
Fix an arbitrary coloring of $[0..u)$. The probability that the generated size-$n$ sequence is encoded by this coloring is equal to the product of $n / \bar{n}$ probabilities that its independently generated size-$\bar{n}$ subsequences are encoded by the coloring restricted to the corresponding blocks, which gives $(1 / \bar{n}^{\Omega(\bar{n})})^{n / \bar{n}} = 1 / \bar{n}^{\Omega(n)}$ (note that, technically, our assumption that gives each probability $1 / \bar{n}^{\Omega(\bar{n})}$ requires the colors in the $i$th block to be from $(i\bar{n}..(i+1)\bar{n}]$ but, clearly, any other colors in the $i$th block make the probability that the corresponding size-$\bar{n}$ subsequence is encoded by this coloring even lower.) The latter, after applying the logarithm, yields the space lower bound $\Omega(n\log\bar{n})$, which is $\Omega(n\log\log\log\frac{u}{n})$ when $\bar{n} = \lfloor(\log\log \frac{u}{n})^{1/3}\rfloor$.

\begin{remark}\label{rem:random}
Using an argument akin to Yao's principle, Assadi et al.~showed that the lower bound for randomized MMPHFs is the same as for deterministic. We repeat their argument for completeness, albeit without their unnecessary graph products etc. 	

A randomized MMPHF has unrestrictedly access to a tape of random bits, which does not take any space. Denote by $X$ the set of all size-$n$ sequences in $[0..u)$. The MMPHF receives a sequence $x \in X$ and a random tape $r$ and encodes $x$ into a memory content $d^r(x) \in \{0,1\}^{*}$. Thus, unlike the deterministic case, the space size depends on $x \in X$ and on the randomness $r$. Naturally, the space occupied by such MMPHF is defined as $d = \max_{x\in X} \mathrm{E}_r [|d^r(x)|]$, where the expectation is for the random $r$. Note that, when the tape $r$ is fixed, the algorithm becomes deterministic: it can be modelled as a function $rank^r \colon [0..u)\times \{0,1\}^* \to [1..n]$ that,  for any memory content $c \in \{0,1\}^*$, defines a coloring of $[0..u)$ into colors $[1..n]$. 

Denote by $P$ a random distribution on $X$ such that any fixed coloring of $[0..u)$ encodes $x \in P$ with probability at most $1/C$. Obviously, $d = \max_{x\in X} \mathrm{E}_r[|d^r(x)|] \ge \mathrm{E}_{x\in P} \mathrm{E}_r[|d^r(x)|] = \mathrm{E}_r \mathrm{E}_{x\in P} [|d^r(x)|]$. By the averaging argument, there is a tape $r^*$ such that $\mathrm{E}_r \mathrm{E}_{x\in P} [|d^r(x)|] \ge \mathrm{E}_{x\in P}[|d^{r^*}(x)|]$. Therefore, $\mathrm{E}_{x\in P}[|d^{r^*}(x)|] \le d$. By Markov's inequality, $\mathrm{Pr}_{x\in P}(|d^{r^*}(x)| \le 2d) \ge \frac{1}{2}$. Denote by $M$ all possible memory contents $d^{r^*}(x)$ for all $x \in P$ such that $|d^{r^*}(x)| \le 2d$. Evidently, we have the lower bound $\frac{1}{2}\log |M| \le d$ and $\mathrm{Pr}_{x\in P}(d^{r^*}(x) \in M) \ge \frac{1}{2}$. Each $c \in M$ determines a coloring of $[0..u)$. Since the probability that a random $x \in P$ is encoded by this coloring is $\frac{1}{C}$, we have $\mathrm{Pr}_{x\in P}(d^{r^*}(x) \in M) \le \frac{|M|}{C}$. Thus, we obtain $|M| \ge \frac{C}{2}$, which implies the space bound $d \ge \Omega(\log C)$, the same as in the deterministic case.
\end{remark}

\section{Random Sequences on Large Universes}
\label{sec:large-universes}

In this section, we always assume that $u = 2^{2^{n^3}}$ and $n \ge 2$.
Our random process generating size-$n$ sequences from $[0..u)$ is essentially a variation of the process by Assadi et al.~\cite{AssadiEtAl}. Unlike the previous sections, it is not precisely a simplification of the arguments from~\cite{AssadiEtAl}, rather a different perspective on them. We first overview main ideas of Assadi et al.~and, then, define our process by modifying them.

\subsection{Definition of the random process}

Let us fix an arbitrary coloring of the range $[0..u)$ into colors $[1..n]$. On a very high level, the random process devised by Assadi et al.\ is as follows: choose $n-1$ lengths $b_2 > \cdots > b_{n}$, then pick uniformly at random $x_1$ from $[0..u)$, then pick uniformly at random $x_2$ from $(x_1 .. x_1{+}b_2)$, then $x_3$ from $(x_2 .. x_2{+}b_3)$, etc. Intuitively, if it is highly likely that at least $\frac{n}2$ of the picks $x_{i} \in (x_{i-1}..x_{i-1}{+}b_{i})$ were such that the fraction of elements with color $i$ in the range $(x_{i-1}..x_{i-1}{+}b_{i})$ is at most $O(\frac{1}n)$, then the probability that the randomly generated sequence $x_1, \ldots, x_n$ is encoded by our fixed coloring is at most $O(\frac{1}n)^{n/2} = \frac{1}{n^{\Omega(n)}}$. Unfortunately, for any $b_2,\ldots,b_{n}$, there are colorings where this is not true. However, as it was shown in~\cite{AssadiEtAl}, when picking $b_2, \ldots, b_{n}$ randomly from a certain distribution such that $b_2 \gg \cdots \gg b_{n}$, one might guarantee that, whenever we encounter a ``dense'' range $(x_{i-1}..x_{i-1}{+}b_{i})$ where at least an $\Omega(\frac{1}{n})$ fraction of colors are $i$, the final range $(x_{n-1} .. x_{n-1}{+}b_{n})$ with very high probability will contain at least a $\frac{2}{n}$ fraction of colors $i$. Therefore, it is highly likely that such ``dense'' ranges appear less than $\frac{n}{2}$ times since otherwise the range $(x_{n-1} .. x_{n-1}{+}b_{n})$ has no room for the color $n$ of element $x_n$ as it already contains $\frac{n}{2}$ colors from $[1..n)$, each occupying a $\frac{2}{n}$ fraction of the range. Hence, with very high probability, the random process generating the size-$n$ sequence will encounter such ``dense'' ranges $(x_{i-1}..x_{i-1}{+}b_{i})$ at most $\frac{n}{2}$ times and at least $\frac{n}{2}$ ranges will contain an $O(\frac{1}{n})$ fraction of the picked color, which leads to the probability $\frac{1}{n^{\Omega(n)}}$ that the generated sequence will be encoded by our fixed coloring.

We alter the outlined scheme introducing a certain ``rigid'' structure into our random process. The range $[0..u)$ is decomposed into a hierarchy of blocks with $L = n^{n^2-n}$ levels (the choice of $L$ is explained below): $[0..u)$ is split into $n^n$ equal blocks, which are called the blocks of level~$1$, each of these blocks is again split into $n^n$ equal blocks, which are the blocks of level~$2$, and so on: for $\ell < L$, each block of level $\ell$ is split into $n^n$ equal blocks of level $\ell+1$. Thus, we have $(n^n)^L = n^{n^{n^2-n+1}}$ blocks on the last level $L$. Observe that $(n^n)^L \le u$ since $\log\log((n^n)^L) = \Theta(n^2\log n) \ll n^3 = \log\log u$. For simplicity, we assume that $(n^n)^L$ divides $u$ (otherwise we could round $u$ to the closest multiple of $(n^n)^L$, ignoring some rightmost elements of $[0..u)$). The length of the last level blocks is set to $u / (n^n)^L$ (their length will not play any role, it is set to this number just to make everything fit into $u$).

\begin{figure}[htb]
	\centering
	\includegraphics[scale=0.87]{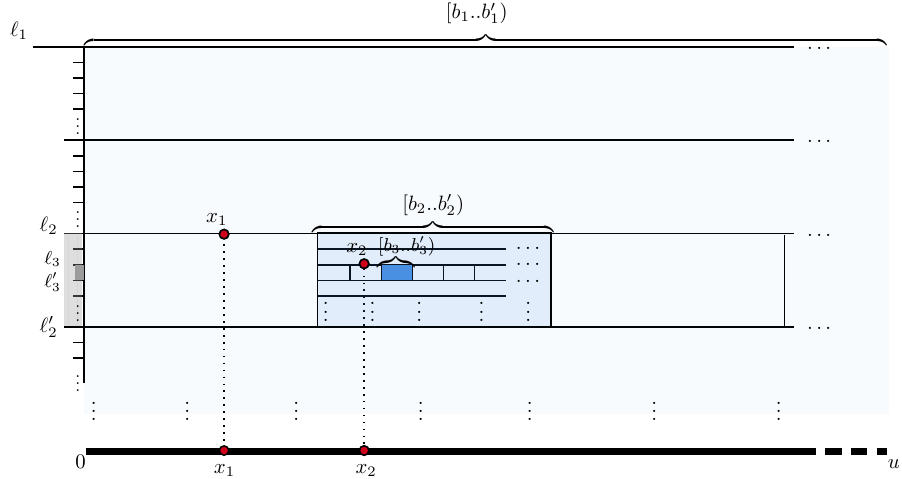}
	\caption{A schematic image of the first intervals $[\ell_1..\ell'_1)$, $[\ell_2..\ell'_2)$, $[\ell_3..\ell'_3)$, the first blocks $[b_1..b'_1)$, $[b_2..b'_2)$, $[b_3..b'_3)$, and the first elements $x_1, x_2$ generated by our process. The set $[0..u)$ is depicted as the line at the bottom. The left vertical ``ruler'' depicts some levels (not all): the larger divisions denote the levels that could be chosen as $\ell_2$ and the smaller divisions could be chosen as $\ell_3$. The intervals $[\ell_2..\ell'_2)$ and $[\ell_3..\ell'_3)$ are painted in two shades of gray. For $i \in [1..3]$, each block $[b_i..b'_i)$ is associated with a rectangle that includes all subblocks of $[b_i..b'_i)$ from levels $[\ell_i..\ell'_i)$; the rectangles are painted in shades of blue; we depict inside the rectangle of $[b_i..b'_i)$ lines corresponding to levels that could be chosen as $\ell_{i+1}$ and we outline contours of blocks from the level $\ell_{i+1}$. The elements $x_1, x_2$ are chosen from $[0..u)$ but it is convenient to draw them also on the lines corresponding to the respective levels $\ell_2$ and $\ell_3$, so it is easier to see that the frist level-$\ell_2$ block to the right of $x_1$ is $[b_2..b'_2)$ and the first level-$\ell_3$ block to the right of $x_2$ is $[b_3..b'_3)$.}\label{fig:process}
\end{figure}

Our random process consists of two parts: first, we pick a sequence of levels $\ell_2 < \cdots < \ell_{n}$; then, we pick the elements $x_1,\ldots,x_n$. The chosen levels $\ell_2,\ldots,\ell_n$ will determine the sizes of $n$ nested blocks from which the elements $x_1,\ldots,x_n$ are sampled. Formally, it is as follows (see Fig.~\ref{fig:process}):
\begin{enumerate}
\item The levels $\ell_2,\ldots,\ell_{n}$ are chosen by consecutively constructing a sequence of nested intervals $[\ell_2..\ell'_2) \supset \cdots \supset [\ell_{n}..\ell'_{n})$: whenever $[\ell_{i}..\ell'_{i})$ is already chosen, for $i \in [1..n)$ (assuming $[\ell_1..\ell'_1) = [0..L)$), we split $[\ell_{i}..\ell'_{i})$ into $n^n$ equal disjoint intervals and pick as $[\ell_{i+1}..\ell'_{i+1})$ one of them uniformly at random, except the first one containing $\ell_{i}$. 
\item The elements $x_1,\ldots,x_n$ are chosen by consecutively constructing a sequence of nested blocks $[b_2..b'_2) \supset \cdots \supset [b_{n}..b'_{n})$ from levels $\ell_2, \ldots, \ell_{n}$, respectively: whenever $[b_{i}..b'_{i})$ is already chosen, for $i \in [1..n)$ (assuming $[b_1..b'_1) = [0..u)$), we pick $x_i$ uniformly at random from the range $[b_{i}..b'_{i}{-}b)$, where $b$ is the block length for level $\ell_{i+1}$ (recall that $b = u / (n^n)^{\ell_{i+1}}$), and choose as $[b_{i+1}..b'_{i+1})$ the block $[kb..(k{+}1)b)$ closest to the right of $x_i$, i.e., $(k{-}1)b \le x_i < kb$; the element $x_n$ is chosen uniformly at random from $[b_{n}..b'_{n})$.
\end{enumerate}

We say that the process \emph{reaches} a block $B$ (respectively, a level $\ell$) on the $i$th stage of recursion if it assigns $[b_i..b'_i) = B$ (respectively, $\ell_i = \ell$) during its work. Note that the length of $[\ell_{i+1}..\ell'_{i+1})$ is $\frac{L}{(n^n)^i}$. Hence, the number $L = n^{n^2-n} = (n^n)^{n-1}$ is large enough to allow the described $n-1$ recursive splits of $[0..L)$. We have $\ell_1 < \ell_2 < \cdots < \ell_n$ (assuming $\ell_1=0$) since, while choosing $[\ell_{i+1}..\ell'_{i+1})$ from $[\ell_{i}..\ell'_{i})$, the process excludes from consideration the first interval $[\ell_i..\ell_i{+}\frac{L}{(n^n)^i})$. Note that, as a byproduct, many levels from $[0..L)$ are not reacheable.

The process can be viewed as a variation on the design of Assadi et al.~restrained to the introduced block structure. Alternatively, it can be viewed as a recursion: for $i \in [1..n)$, it takes as its input an interval $[\ell_{i}..\ell'_{i})$ and a block $[b_{i}..b'_{i})$ from level $\ell_{i}$ (assuming  $[\ell_1..\ell'_1) = [0..L)$ and $[b_1..b'_1) = [0..u)$), chooses randomly $\ell_{i+1}$ from a set of $n^n-1$ evenly spaced levels in $(\ell_{i}..\ell'_{i})$, then picks $x_i$, and invokes the recursion to generate the elements $x_{i+1}, \ldots, x_n$ inside the closest level-$\ell_{i+1}$ block to the right of $x_i$, setting $[\ell_{i+1}..\ell'_{i+1})$ as the next interval of levels. In this view, the process is not split into two parts and it constructs the levels and blocks simultaneously, which is possible since the levels are chosen independently of the blocks.

\begin{figure}
	\centering
	\includegraphics[scale=0.87]{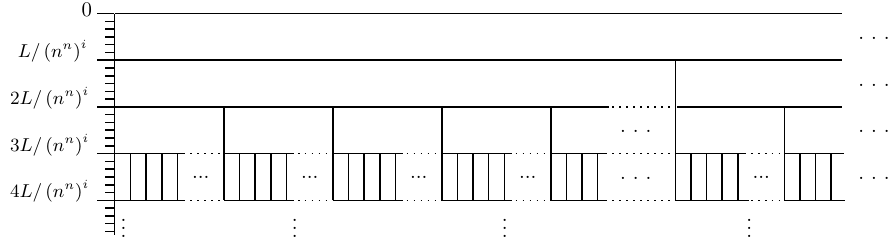}
	\caption{A schematic partition of all blocks into disjoint subsets for a fixed $i \in [1..n)$.}\label{fig:partition}
\end{figure}

The nestedness of the intervals $[\ell_{i}..\ell'_{i})$ and the blocks $[b_{i}..b'_{i})$ implies that, whenever the process reaches a block $[b_{n}..b'_{n})$ on level $\ell_n$, we can uniquely determine the sequence of intervals $[\ell_{2}..\ell'_{2}), \ldots, [\ell_{n}..\ell'_{n})$ and blocks $[b_{2}..b'_{2}), \ldots, [b_{n}..b'_{n})$ that were traversed. It is instructive to keep in mind the following view (Fig.~\ref{fig:partition}). Fix $i\in [1..n)$. Split $[0..L)$ into $(n^n)^i$ equal disjoint intervals: $I = \{[k\frac{L}{(n^n)^i}..(k+1)\frac{L}{(n^n)^i})\}_{k\in [0..(n^n)^i)}$. The set of all blocks can be partitioned into disjoint subsets as follows: each subset is determined by an interval $[\ell..\ell')$ from $I$ and a level-$\ell$ block $[b..b')$ and consists of all subblocks of $[b..b')$ from levels $[\ell..\ell')$ (including $[b..b')$ itself). Then, the $(i+1)$th recursive invocation of our process (which chooses $x_{i+1}$) necessarily takes as its input an interval $[\ell..\ell')$ from $I$ and a level-$\ell$ block $[b..b')$, and subsequently it can reach only subblocks from the corresponding set in the partition. Note, however, that not all subblocks are reachable since, first, some levels are unreachable, as was noted above, and, second, the process ignores the leftmost subblock of its current block when it chooses the next block (since this subblock is not located to the right of any element $x$ in the block). For the same reason, not all intervals $[\ell..\ell') \in I$ and level-$\ell$ blocks $[b..b')$ might appear as inputs of the recursion.

\subsection{Analysis of the random process} 

Fix an arbitrary coloring of $[0..u)$ into colors $[1..n]$, which will be used untill the end of this section. We call a subset of $[0..u)$ \emph{dense} for color $i$ if at least a $\frac{2}{n}$ fraction of its elements have color $i$; we call it \emph{sparse} otherwise. The color is not specified if it is clear from the context. 

\begin{figure}[htb]
	\centering
	\includegraphics[scale=0.87]{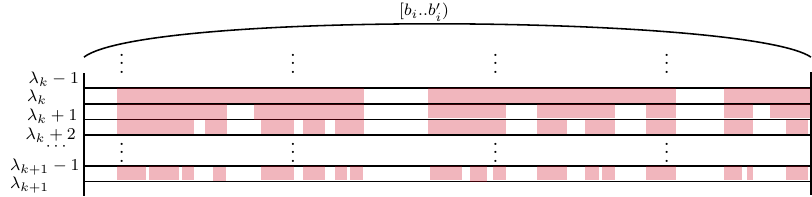}
	\caption{The lines depict consecutive levels $[\lambda_k-1..\lambda_{k+1}]$ inside a block $[b_i..b'_i)$; we assume that $[\ell_{i+1}..\ell'_{i+1}) = [\lambda_k..\lambda_{k+1})$. The red regions denote the dense sets $D_{\ell}$, for $\ell \in [\lambda_k-1..\lambda_{k+1})$. The image is supposed to show the case when each such $D_{\ell}$ takes a large portion of $D_{\lambda_k-1}$ , so that $D_{\lambda_k-1}$ might (approximately) serve as our ``inherently dense'' set $\bar{D}_k$ for level $\lambda_k$ in the block.}\label{fig:dense}
\end{figure}

\subparagraph{Analysis plan.}
For each $i \in [1..n)$ and each block $[b_i..b'_i)$ that might appear on the $i$th stage of our recursion, we show that whenever the process reaches $[b_i..b'_i)$, the level $\ell_{i+1}$ it randomly chooses (among $n^n-1$ choices) admits, with high probability $1 - \frac{1}{n^{\Omega(n)}}$, a partition of all level-$\ell_{i+1}$ blocks inside $[b_i..b'_i)$ into two disjoint families (see Fig.~\ref{fig:dense}): a ``sparse'' set $\bar{S}$, whose union is sparse for color $i$, and an ``inherently dense'' set $\bar{D}$ of blocks, each of which is dense for color $i$ and almost all subblocks of $\bar{D}$ on each subsequent level $\ell \in [\ell_{i+1}..\ell'_{i+1})$ are dense for $i$ too, where ``almost all'' means that only a $\frac{1}{n^{\Omega(n)}}$ fraction of level-$\ell$ subblocks of $\bar{D}$ might be sparse. Thus, whenever the process chooses $x_i$ from the ``inherently dense'' set on an $i$th stage, it will with high probability $1 - \frac{1}{n^{\Omega(n)}}$ end up inside a block $[b_n..b'_n)$ dense for color $i$, where it picks $x_n$ in the end. Therefore, to have a room for one element $x_n$ with color $n$, such hits into ``inherently dense'' sets could happen on less than $\frac{n}{2}$ different stages $i$ in the recursion, with high probability. We deduce from this, like in the scheme from~\cite{AssadiEtAl} outlined above, that at least $\frac{n}{2}$ stages $i$ of the recursion pick $x_i$ from the sparse sets $\bar{S}$, with high probability, which allows us to estimate by $O(\frac{1}{n})^{\frac{n}{2}} = \frac{1}{n^{\Omega(n)}}$ the probability that the generated sequence is correctly colored in our fixed coloring. Now let us formalize this.

\subparagraph{\boldmath Constructing $\bar{S}$ and $\bar{D}$.}
Fix $i\in [1..n)$. Let $H = [\ell_{i}..\ell'_{i})$ and $B = [b_{i}..b'_{i})$ be, respectively, an interval of levels and a level-$\ell_{i}$ block that could be reached by our process on the $i$th stage of recursion (assuming that $[\ell_1..\ell'_1) = [0..L)$ and $[b_1..b'_1) = [0..u)$).
For $\ell \in H$, denote by $S_\ell$ the union of all sparse blocks from levels $[\ell_{i}..\ell]$ that are subsets of $B$. Due to the nestedness of blocks, $S_\ell$ is equal to the union of $S_{\ell-1}$ (assuming $S_{\ell-1} = \emptyset$ for $\ell = \ell_{i}$) and all sparse level-$\ell$ blocks that are subsets of $B$ disjoint with $S_{\ell-1}$. Obviously, the set $S_\ell$ is sparse. Denote $D_\ell = B \setminus S_\ell$, the complement of $S_\ell$, which is equal to the union of all dense level-$\ell$ blocks that are subsets of $B$ disjoint with $S_\ell$ (note that some dense level-$\ell$ blocks could be subsets of $S_\ell$ if they were covered by larger sparse blocks). Consult Figure~\ref{fig:dense} in what follows.

Denote $\lambda_0 = \ell_{i}$ and $\lambda_k = \lambda_{k-1} + \frac{|H|}{n^n}$, for $k \in [1..n^n]$. The intervals $[\lambda_k..\lambda_{k+1})$, for all $k \in [1..n^n)$, are all possible choices for the random interval $[\ell_{i+1}..\ell'_{i+1})$. To choose the interval is to choose $k \in [1..n^n)$. We slightly relax the scheme outlined in the plan: for each of the choices $k \in [1..n^n)$, we define a set $\bar{S}_k$ that will contain a $\frac{2}{n}+\frac{1}{2^{n/8}}$ fraction of colors $i$, so it might be not precisely sparse as in the plan. We call sets with this fraction of a given color \emph{almost sparse}. If $B$ itself is almost sparse, we define $\bar{S}_k = B$ and $\bar{D}_k = \emptyset$, for all $k\in [1..n^n)$. Otherwise, i.e., when $B$ is not almost sparse, we are to prove that, for a randomly chosen level $\ell_{i+1}$, with high probability $1 - \frac{1}{n^{\Omega(n)}}$ not only the blocks composing $D_{\ell_{i+1}}$ are dense but also most of their subblocks on levels $\ell \in [\ell_{i+1}..\ell'_{i+1})$ are dense for color $i$. The sets $\bar{D}_k$ will be constructed using the sets $D_{\ell_{i+1}}$ with this property. So, assume that $B$ is not almost sparse.

Let $D_{\ell_{i}-1} = B$. Since the sets $D_\ell$ are nested (i.e., $D_{\ell-1} \supseteq D_{\ell}$), the ``fraction of space'' which any $D_{\ell+\delta}$ occupies inside any $D_{\ell}$ is $\frac{|D_{\ell+\delta}|}{|D_{\ell}|}$ (a number between $0$ and $1$). Therefore, for $k \in [0..n^n)$, the fraction of space which each of the sets $D_{\lambda_k}, D_{\lambda_k+1}, \ldots, D_{\lambda_{k+1}-1}$ occupies inside the set $D_{\lambda_k-1}$ is at least $q_{\lambda_k} = \frac{|D_{\lambda_{k+1}-1}|}{|D_{\lambda_{k}-1}|}$. 
Observe that $\prod_{k=0}^{n^n-1} q_{\lambda_k}$ is equal to the fraction of space that $D_{\ell'_i-1}$ occupies in $B$. This product is greater than $\frac{1}{2^{n/8}}$ since otherwise the fraction of colors $i$ in $B = S_{\ell'_i-1} \cup D_{\ell'_i-1}$ is at most $\frac{2}n + \frac{1}{2^{n/8}}$, contrary to our assumption that $B$ is not almost sparse. Hence, $\prod_{k=0}^{n^n-1} q_{\lambda_k} > \frac{1}{2^{n/8}}$. The product with this many (namely $n^n$) factors cannot contain many even mildly small values $q_{\lambda_k}$ provided the result is as large as $\frac{1}{2^{n/8}}$: indeed, if we have at least $n^{n/2}$ factors that are at most $1 - \frac{1}{n^{n/4}}$, then we already obtain $(1 - \frac{1}{n^{n/4}})^{n^{n/2}} = ((1 - \frac{1}{n^{n/4}})^{n^{n/4}})^{n^{n/4}} = O(\frac{1}{e^{n^{n/4}}})$, much smaller than $\frac{1}{2^{n/8}}$. Thus, less than $n^{n/2}$ numbers $q_{\lambda_1}, \ldots, q_{\lambda_{n^n-1}}$ can be less than $1 - \frac{1}{n^{n/4}}$ and, for most $k\in [1..n^n)$, we have $q_{\lambda_k} \ge 1 - \frac{1}{n^{n/4}}$. Therefore, the probability that $q_{\ell_{i+1}} < 1 - \frac{1}{n^{n/4}}$, where $\ell_{i+1}$ is chosen uniformly at random among the levels $\lambda_1, \ldots, \lambda_{n^n-1}$, is at most $\frac{n^{n/2}}{n^n-1} = O(\frac{1}{n^{n/2}})$. For $k \in [1..n^n)$, we call the level $\lambda_k$ \emph{abnormal for $B$} if $q_{\lambda_k} < 1 - \frac{1}{n^{n/4}}$, and \emph{normal} otherwise.

\begin{figure}[htb]
	\centering
	\includegraphics[scale=0.87]{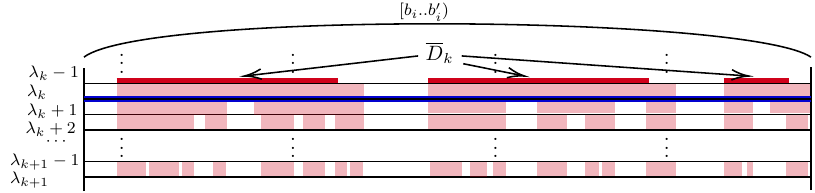}
	\caption{The lines depict consecutive levels $[\lambda_k-1..\lambda_{k+1}]$ inside a block $[b_i..b'_i)$. The level $\lambda_k$ is emphasized by the blue color. The red region under line representing level $\ell$ depicts $D_{\ell}$.  The set $D_{\lambda_k-1}$ consists of three maximal intervals; accordingly, $\bar{D}_k$ is drawn as three thick red lines over $D_{\lambda_k-1}$ (the gap to the right of each line represents the lacking rightmost block from level $\lambda_k$).}\label{fig:inherent}
\end{figure}

Let us define, for each normal level $\lambda_k$ with $k \in [1..n^n)$, a partition of $B$ into an almost sparse set $\bar{S}_k$ and an ``inherently dense'' set $\bar{D}_k$ that were announced above.
One might suggest that $\bar{D}_k$ can be defined as $D_{\lambda_k}$. Indeed, it seems to have the alluded property. However, observe that even when the randomly chosen $x_{i+1}$ ``hits'' $D_{\lambda_k}$, the block $[b_{i+1}..b'_{i+1})$, which is the first level-$\lambda_k$ block to the right of $x_{i+1}$, might not be a subset of $D_{\lambda_k}$. So, there is no ``inheritance'' after hitting $D_{\lambda_k}$. Our trick is to define $\bar{D}_k$ as $D_{\lambda_k-1}$ minus the rightmost blocks from level $\lambda_k$ on each maximal interval in $D_{\lambda_k-1}$ (see Fig.~\ref{fig:inherent}). This trick is the reason why we defined the number $q_{\lambda_k}$ as the fraction of space that $D_{\lambda_{k+1}-1}$ occupies in $D_{\lambda_k-1}$, not in $D_{\lambda_k}$. To formalize this, let us decompose $D_{\lambda_k-1}$ into maximal intervals: $D_{\lambda_k-1} = [d_1..d'_1) \cup \cdots \cup [d_t..d'_t)$, where $d'_j < d_{j+1}$ for $j\in [1..t)$. All the intervals are aligned on block boundaries for blocks from both levels $\lambda_k-1$ and $\lambda_k$. Denote by $b$ the block length on level $\lambda_k$. Since the block length on level $\lambda_k-1$ is $n^n b$, the length of each interval is at least $n^n b$ and the distance between the intervals is at least $n^n b$. If $D_{\lambda_k-1} = B$, define $\bar{D}_k = B$ and $\bar{S}_k = \emptyset$; otherwise, define $\bar{D}_k =  [d_1..d'_1{-}b) \cup \cdots \cup [d_t..d'_t{-}b)$ and $\bar{S}_k = B \setminus \bar{D}_k$. Hence, $\bar{S}_k$ is $S_{\lambda_k-1}$ plus $t$ blocks of size $b$. Since $\bar{S}_k$ contains at least $t-1$ disjoint intervals each with length at least $n^n b$ (those intervals are the distances between the intervals of $D_{\lambda_k-1}$), these added blocks constitute at most a $2\frac{b}{n^n b} = \frac{2}{n^{n}}$ fraction of the size of $S_{\lambda_k-1}$. Hence, the fraction of colors $i$ in $\bar{S}_k$ is at most $\frac{2}{n} + \frac{2}{n^{n}}$, which is enough for $\bar{S}_k$ to be almost sparse.

For each abnormal level $\lambda_k$ in the block $B$, we call all subblocks of $B$ from levels $[\lambda_k..\lambda_{k+1})$ \emph{abnormal}; for each normal level $\lambda_k$ with $k \in [1..n^n)$, we call all sparse subblocks of $D_{\lambda_k-1}$ from levels $[\lambda_k..\lambda_{k+1})$ \emph{abnormal}. 
Thus, for the fixed $i \in [1..n)$ and the fixed level-$\ell_i$ block $B$ that could be reached by our process on the $i$th stage of recursion, we have defined abnormal levels, abnormal blocks, and the sets $\bar{D}_k$ and $\bar{S}_k$, for all normal levels $\lambda_k$ that could be chosen as $\ell_{i+1}$ for the $(i+1)$th stage of recursion. Analogously, for all $i \in [1..n)$, we define abnormal levels, abnormal blocks, and sets $\bar{D}_k$ and $\bar{S}_k$ for all blocks $B$ that could possibly be reached by our process on the $i$th stage of recursion. All non-abnormal blocks are called \emph{normal}.

\begin{remark}\label{rem:normal}
The crucial observation about the normal blocks is as follows: if the last block $[b_n..b'_n)$ reached by our process on the $n$th stage is normal, then, for each $i \in [1..n)$, this last block can be sparse for color $i$ only if the element $x_i$ was chosen by the process from an almost sparse set $\bar{S}_k$ defined in the corresponding block $B = [b_i..b'_i)$ for level $\ell_{i+1} = \lambda_k$ (this level  $\lambda_k$ must be normal for $B$ since the last block is normal). This behaviour corresponds to our expectations outlined in the beginning of this section: whenever $x_i$ ``hits'' an ``inherently dense'' set $\bar{D}$ corresponding to the level $\ell_{i+1}$ in the block $B$, it is guaranteed that the last block $[b_n..b'_n)$ will be dense for color $i$, provided this last block is normal. The idea is that the abnormal blocks are unlikely to appear as last blocks in the process and we will be able to restrict our attention only to normal blocks.
\end{remark}

\subparagraph{Probability to end up in an abnormal block.}
For $i \in [1..n)$, denote by $\mathcal{B}_i$ all blocks that could possibly be reached by the process on the $i$th stage of recursion. Let us estimate the probability that the last block $[b_n..b'_n)$ produced by our process is abnormal as follows:
$$
\sum_{i=1}^{n-1} \sum_{B\in\mathcal{B}_i} \Pr\left(\begin{matrix}\text{the process}\\ \text{reaches block }B\end{matrix}\right) \cdot \Pr\left(\begin{matrix}\text{}[b_n..b'_n)\text{ ends up being abnormal}\\ \text{after the process reaches }B\end{matrix}\right).
$$
For each fixed $i$, the second sum is through disjoint events ``the process reaches block $B$'', for $B\in\mathcal{B}_i$; thus, the sum of the probabilities for these events (with the fixed $i$) is $1$. Therefore, if we prove that, for the fixed $i$ and any fixed $B\in\mathcal{B}_i$, the probability of the event ``$[b_n..b'_n)$ ends up being abnormal after the process reaches $B$'' is at most $O(\frac{1}{n^{n/4}})$, then the total sum is bounded as follows:
\begin{equation}\label{eq:total-abnorm}
\sum_{i=1}^{n-1} \sum_{B\in\mathcal{B}_i} \Pr\left(\begin{matrix}\text{the process}\\ \text{reaches block }B\end{matrix}\right)\cdot O\left(\frac{1}{n^{n/4}}\right) = \sum_{i=1}^{n-1} O\left(\frac{1}{n^{n/4}}\right) = O\left(\frac{n}{n^{n/4}}\right) \le O\left(\frac{1}{n^{n/8}}\right).
\end{equation}

Suppose that the process reaches a block $B$ on the $i$th stage of recursion. Case (i): it may end up in an abnormal block $[b_n..b'_n)$ if $\ell_{i+1}$ happens to be an abnormal level. The probability of this is $O(\frac{1}{n^{n/2}})$ since, as was shown, less than $n^{n/2}$ of $n^n-1$ possible choices for the level $\ell_{i+1}$ are abnormal and $\ell_{i+1}$ is chosen uniformly at random. Case (ii): the probability that $[b_n..b'_n)$ ends up being abnormal while $\ell_{i+1}$ is normal can be estimated as follows (the sum is taken over all normal levels among $\lambda_1,\ldots,\lambda_{n^n-1}$ defined for our fixed block $B$):
\begin{equation}\label{eq:abnorm}
\sum_{\begin{matrix}\ell\in [\lambda_k..\lambda_{k+1})\\\text{ for normal }\lambda_k\end{matrix}} \Pr(\ell_{n} = \ell)\cdot \Pr\left(\begin{matrix}[b_n..b'_n)\text{ is abnormal}\\ \text{block of level }\ell\end{matrix}\right).
\end{equation}
Since the events ``$\ell_n = \ell$'' are disjoint, the sum of $\Pr(\ell_n=\ell)$ is $1$ (note that $\Pr(\ell_n = \ell) = 0$, for $\ell$ unreachable on the $n$th stage). Therefore, if, for any normal $\lambda_k$ and $\ell\in [\lambda_k..\lambda_{k+1})$, we estimate by $O(\frac{1}{n^{n/4}})$ the probability that $[b_n..b'_n)$ ends up being an abnormal subblock of $B$ on level $\ell$, then the sum~(\ref{eq:abnorm}) is upperbounded by $O(\frac{1}{n^{n/4}})$. 

Our random process is designed in such a way that, for any $\ell \in [\lambda_k..\lambda_{k+1})$, it reaches on the $n$th stage any reacheable level-$\ell$ subblock of $B$ with equal probability.
Since, for any normal level $\lambda_k$, we have $q_{\lambda_k} \ge 1 - \frac{1}{n^{n/4}}$, the fraction of abnormal subblocks of $B$ on any level $\ell\in [\lambda_k..\lambda_{k+1})$ is at most $\frac{1}{n^{n/4}}$. However, not all level-$\ell$ subblocks are reachable since the process always ignores the leftmost block when it chooses the block for the next stage (for instance, when we uniformly at random pick one of level-$\lambda_k$ subblocks of $B$ for the recursion to stage $i+1$, we cannot choose the leftmost subblock, as it is not located to the right of any $x_i\in B$). Since the number of subblocks for the choice is always at least $n^n$, the dismissed leftmost subblock renders unreachable at most a $\frac{1}{n^n}$ fraction of level-$\ell$ subblocks of $B$. Such dismissals happen for each of the stages $i,i+1,\ldots,n-1$. Hence, the fraction of unreachable level-$\ell$ subblocks of $B$ is at most $\frac{n}{n^n}$. Consequently, the probability that one of the (equally probable) reachable level-$\ell$ subblocks of $B$ is abnormal is at most $\frac{1}{n^{n/4}} / (1 - \frac{n}{n^n}) = O(\frac{1}{n^{n/4}})$.

Adding cases~(i) and~(ii), we obtain the probability $O(\frac{1}{n^{n/2}} + \frac{1}{n^{n/4}}) = O(\frac{1}{n^{n/4}})$ to reach an abnormal block after reaching the block $B$, which, due to the sum (\ref{eq:total-abnorm}), leads to the total probability $O(\frac{1}{n^{n/8}})$ that the last block $[b_n..b'_n)$ in the process ends up being abnormal.

\subparagraph{Probability of correct coloring.}
Now we estimate the probability that the increasing size-$n$ sequence $x_1,\ldots,x_n$ generated by our process is correctly encoded by our fixed coloring of $[0..u)$, i.e., the color of $x_i$ is $i$, for each $i \in [1..n]$. Suppose that the process generates a sequence $x_1,\ldots,x_n$ and, during its work, reaches levels $\ell_1,\ldots,\ell_n$ and blocks $[b_1..b'_1),\ldots,[b_n..b'_n)$ such that the block $[b_n..b'_n)$ is normal. For each $i\in[1..n)$, let $[b_i..b'_i) = \bar{S}^i \cup \bar{D}^i$ be the described above partition of level-$\ell_{i+1}$ subblocks of $[b_i..b'_i)$ into an almost sparse set $\bar{S}^i$ and an ``inherently dense'' set $\bar{D}^i$ (we use the upper indices to avoid confusion with the notation $\bar{S}_k, \bar{D}_k$ used for the partitions on different levels $\ell_{i+1}$, not on different stages as we do now). Then, the sequence $x_1,\ldots,x_n$ might be correctly encoded by our fixed coloring of $[0..u)$ only if we had $x_i \in \bar{S}^i$, for at least $\frac{n}{2}$ stages $i \in [1..n)$, since otherwise the whole block $[b_n..b'_n)$ will be dense for at least $\frac{n}{2}$ different colors from $[1..n)$, lacking a room for color $n$ to paint $x_n \in [b_n..b'_n)$ (here we rely on Remark~\ref{rem:normal} about normal blocks above).

According to this observation, the probability that the generated sequence is correctly encoded can be estimated by the sum of the following numbers (a) and (b): 
\begin{alphaenumerate}
\item the probability that $[b_n..b'_n)$ is abnormal, 
\item the probability that the sequence $x_1,\ldots,x_n$ is correctly encoded, subject to the condition that $x_i \in \bar{S}^i$, for at least $\frac{n}{2}$ stages $i \in [1..n)$ of the process that produced $x_1,\ldots,x_n$. 
\end{alphaenumerate}
This estimation covers all possible generated sequences except those for which the process ended up in a normal block $[b_n..b'_n)$ but less than $\frac{n}{2}$ stages $i \in [1..n)$ had $x_i \in \bar{S}^i$; but this case can be dismissed since the probability for such sequences to be correctly encoded is zero, as was observed above (they have no room for color $n$ in  $[b_n..b'_n)$). We have already deduced that (a) is $O(\frac{1}{n^{n/8}})$. It remains to estimate (b) as $\frac{1}{n^{\Omega(n)}}$.

Fix $M \subseteq [1..n)$ such that $|M|\ge \frac{n}{2}$. Let us estimate the probability $p_M$ that the generated sequence $x_1,\ldots,x_n$ is correctly encoded and satisfies the following condition: $x_i \in \bar{S}^i$, for $i \in M$, and $x_i \in \bar{D}^i$, for $i\not\in M$, where $i \in [1..n)$ are the stages of the process that produced $x_1,\ldots,x_n$. We then can upperbound (b) by $\sum_M p_M$, where the sum is through all $M \subseteq [1..n)$ such that $|M|\ge \frac{n}{2}$. If we show that $p_M < \frac{1}{n^{\Omega(n)}}$, then this sum can be bounded by $\frac{2^n}{n^{\Omega(n)}}$, which is equal to $\frac{1}{n^{\Omega(n)}}$. Indeed, for a fixed $M$ and $i \in M$, when the process reaches the $i$th stage of recursion, the probability that the randomly chosen $x_i$ belongs the set $\bar{S}^i$ and has color $i$ is at most $\frac{2}{n} + \frac{1}{2^{n/8}} \le \frac{3}{n}$ since the set $\bar{S}^i$ is almost sparse (note that the probability is zero if $\bar{S}^i = \emptyset$). Then, the probability that $x_i$ belongs to $\bar{S}^i$ and has color $i$, for all $i \in M$, is at most $(\frac{3}{n})^{|M|} \le (\frac{3}{n})^{n/2} = \frac{1}{n^{\Omega(n)}}$.



\bibliography{mmphf-bound}

\begin{thebibliography}{10}

\bibitem{AssadiEtAl}
S.~Assadi, M.~Farach-Colton, and W.~Kuszmaul.
\newblock Tight bounds for monotone minimal perfect hashing.
\newblock In {\em Proc. Annual ACM-SIAM Symposium on Discrete Algorithms
  (SODA)}, pages 456--476. SIAM, 2023.
\newblock \href {https://doi.org/10.1137/1.9781611977554.ch2}
  {\path{doi:10.1137/1.9781611977554.ch2}}.

\bibitem{Belazzougui2}
D.~Belazzougui.
\newblock Linear time construction of compressed text indices in compact space.
\newblock In {\em Proceedings of the forty-sixth Annual ACM Symposium on Theory
  of Computing}, pages 148--193, 2014.
\newblock \href {https://doi.org/10.1145/2591796.2591885}
  {\path{doi:10.1145/2591796.2591885}}.

\bibitem{BelazzouguiBoldiPaghVigna}
D.~Belazzougui, P.~Boldi, R.~Pagh, and S.~Vigna.
\newblock Monotone minimal perfect hashing: searching a sorted table with
  {O}(1) accesses.
\newblock In {\em Proc. SODA}, pages 785--794. SIAM, 2009.
\newblock \href {https://doi.org/10.1137/1.9781611973068.86}
  {\path{doi:10.1137/1.9781611973068.86}}.

\bibitem{BelazzouguiBotelhoDietzfelbinger}
D.~Belazzougui, F.~C. Botelho, and M.~Dietzfelbinger.
\newblock Hash, displace, and compress.
\newblock In {\em European Symposium on Algorithms}, pages 682--693. Springer,
  2009.
\newblock \href {https://doi.org/10.1007/978-3-642-04128-0_61}
  {\path{doi:10.1007/978-3-642-04128-0_61}}.

\bibitem{BelazzouguiCunialKarkkainenMakinen}
D.~Belazzougui, F.~Cunial, J.~K{\"a}rkk{\"a}inen, and V.~M{\"a}kinen.
\newblock Linear-time string indexing and analysis in small space.
\newblock {\em ACM Transactions on Algorithms (TALG)}, 16(2):1--54, 2020.
\newblock \href {https://doi.org/10.1145/3381417} {\path{doi:10.1145/3381417}}.

\bibitem{BelazzouguiNavarro}
D.~Belazzougui and G.~Navarro.
\newblock Alphabet-independent compressed text indexing.
\newblock {\em ACM Transactions on Algorithms (TALG)}, 10(4):1--19, 2014.
\newblock \href {https://doi.org/10.1145/2635816} {\path{doi:10.1145/2635816}}.

\bibitem{BelazzouguiNavarro2}
D.~Belazzougui and G.~Navarro.
\newblock Optimal lower and upper bounds for representing sequences.
\newblock {\em ACM Transactions on Algorithms (TALG)}, 11(4):1--21, 2015.
\newblock \href {https://doi.org/10.1145/2629339} {\path{doi:10.1145/2629339}}.

\bibitem{Clark}
D.~Clark.
\newblock {\em Compact pat trees}.
\newblock PhD thesis, University of Waterloo, 1997.

\bibitem{CliffordEtAl}
R.~Clifford, A.~Fontaine, E.~Porat, B.~Sach, and T.~Starikovskaya.
\newblock Dictionary matching in a stream.
\newblock In {\em Proc. ESA}, volume 9294 of {\em LNCS}, pages 361--372.
  Springer, 2015.
\newblock \href {https://doi.org/10.1007/978-3-662-48350-3_31}
  {\path{doi:10.1007/978-3-662-48350-3_31}}.

\bibitem{CoverThomas}
T.~M. Cover and J.~A. Thomas.
\newblock Information theory and statistics.
\newblock {\em Elements of Information Theory}, 1(1):279--335, 1991.
\newblock \href {https://doi.org/10.1002/0471200611}
  {\path{doi:10.1002/0471200611}}.

\bibitem{FredmanKomlos}
M.~L. Fredman and J.~Koml{\'o}s.
\newblock On the size of separating systems and families of perfect hash
  functions.
\newblock {\em SIAM Journal on Algebraic Discrete Methods}, 5(1):61--68, 1984.
\newblock \href {https://doi.org/10.1137/0605009} {\path{doi:10.1137/0605009}}.

\bibitem{FredmanKomlosSzemeredi}
M.~L. Fredman, J.~Koml{\'o}s, and E.~Szemer{\'e}di.
\newblock Storing a sparse table with {O}(1) worst case access time.
\newblock {\em Journal of the ACM}, 31(3):538--544, 1984.
\newblock \href {https://doi.org/10.1145/828.1884}
  {\path{doi:10.1145/828.1884}}.

\bibitem{GagieNavarroPrezza2}
T.~Gagie, G.~Navarro, and B.~Prezza.
\newblock Fully functional suffix trees and optimal text searching in bwt-runs
  bounded space.
\newblock {\em Journal of the ACM (JACM)}, 67(1):1--54, 2020.
\newblock \href {https://doi.org/10.1145/3375890} {\path{doi:10.1145/3375890}}.

\bibitem{GrossiOrlandiRaman}
R.~Grossi, A.~Orlandi, and R.~Raman.
\newblock Optimal trade-offs for succinct string indexes.
\newblock In {\em Automata, Languages and Programming: 37th International
  Colloquium, ICALP 2010, Bordeaux, France, July 6-10, 2010, Proceedings, Part
  I 37}, pages 678--689. Springer, 2010.
\newblock \href {https://doi.org/10.1007/978-3-642-14165-2_57}
  {\path{doi:10.1007/978-3-642-14165-2_57}}.

\bibitem{Jacobson}
G.~Jacobson.
\newblock Space-efficient static trees and graphs.
\newblock In {\em Proc. 30th Annual Symposium on Foundations of Computer
  Science (FOCS)}, pages 549--554. IEEE, 1989.
\newblock \href {https://doi.org/10.1109/SFCS.1989.63533}
  {\path{doi:10.1109/SFCS.1989.63533}}.

\bibitem{Mehlhorn}
K.~Mehlhorn.
\newblock On the program size of perfect and universal hash functions.
\newblock In {\em 23rd Annual Symposium on Foundations of Computer Science
  (SFCS 1982)}, pages 170--175. IEEE, 1982.
\newblock \href {https://doi.org/10.1109/SFCS.1982.80}
  {\path{doi:10.1109/SFCS.1982.80}}.

\bibitem{Radhakrishnan}
J.~Radhakrishnan.
\newblock Improved bounds for covering complete uniform hypergraphs.
\newblock {\em Information Processing Letters}, 41(4):203--207, 1992.
\newblock \href {https://doi.org/10.1016/0020-0190(92)90181-T}
  {\path{doi:10.1016/0020-0190(92)90181-T}}.

\end{thebibliography}

\end{document}